\documentclass[11pt]{article}
\pdfoutput=1
\usepackage{jheppub}
\usepackage{graphicx}
\usepackage{amssymb}
\usepackage{amsmath}
\usepackage{calc}
\usepackage{slashed}
\usepackage{caption}
\usepackage{xcolor}
\usepackage{dsfont}
\usepackage{verbatim}
\usepackage{enumitem}
\usepackage{subfig}

\newcommand\beq{\begin{equation}}
\newcommand\eeq{\end{equation}}

\title{Phases of Flavor Broken QCD$_3$}

\preprint{\today}

\author[a]{Andrew Baumgartner}

\affiliation[a]{Department of Physics, University of Washington, Seattle, WA, 98195-1560, USA}

\abstract{We map out the phase diagram of QCD$_3$ with a product flavor group of the form $U(f)\times U(F)$. We find interesting structures emerge when $f+F>k$ depending on the relative sizes of $f$,$F$ and $k$. In particular, there exists phase transitions in which a Grassmannian phase will disappear and reappear in a different part of the phase diagram.}

\begin{document}
\maketitle

\section{Introduction}
Recently there has been a renewed interest in the phases of 2+1 dimensional gauge theories. This flurry of work was largely inspired by condensed matter systems with emergent gauge fields \cite{Son:2015xqa}, large N limits of vector-like gauge theories \cite{Minwalla:2015sca,Giombi:2011kc,Aharony:2011jz,Aharony:2012nh,Jain:2014nza} higher spin gravity \cite{Giombi:2009wh,Giombi:2012ms} and supersymmetric dualities \cite{Kachru:2016aon,Gur-Ari:2015pca}. These efforts culminated in a set of infra-red dualities between non-supersymmetric Chern-Simons gauge theories with fundamental matter \cite{Aharony:2015mjs,Hsin:2016blu}. One such duality is
\begin{equation} \label{eq:aharony}
SU(N)_{-k+\frac{N_f}{2}} \text{ with } N_f \, \, \psi \,\,\,\,\leftrightarrow\,\,\,\, U(k)_N \text{ with } N_f \,\, \phi
\end{equation} 
which relates QCD$_3$ with massless quarks to a Wilson-Fisher scalar. A rigorous proof of this duality is still out of reach, but it has undergone some stringent consistency checks since it was first conjectured. Examples include explicit large $N$ calculations \cite{Minwalla:2015sca,Giombi:2011kc,Aharony:2011jz,Aharony:2012nh,Jain:2014nza}, anomaly matching \cite{Benini:2017dus}, consistency of boundary conditions and anomaly inflow \cite{Aitken:2018joi,Aitken:2017nfd}, holographic embeddings \cite{Jensen:2017xbs}, and lattice constructions \cite{Chen:2018vmz,Son:2018zja,Chen:2017lkr} to name a few. One of the simplest consistency checks one can perform is the matching of massive phases on either side of the duality. It is not difficult to check that under a mapping of the form $\bar{\psi}\psi\leftrightarrow -|\phi|^2$ the phases match. However if $N_f>k$, a negative mass deformation for the scalar will completely break the gauge group and the low energy limit is described by a non-linear sigma model (NLSM). No such NLSM is visible on the fermionic side. This observation, along with some explicit large $N_f$ calculations \cite{Appelquist:1989tc}, lead to the  ``flavor-bound": $N_f\le k$. The duality \eqref{eq:aharony} is only valid if the flavor bound is satisfied. Otherwise, QCD$_3$ can not be described by a scalar dual. 

\begin{figure}[t]
\begin{centering}
\includegraphics[scale=0.50]{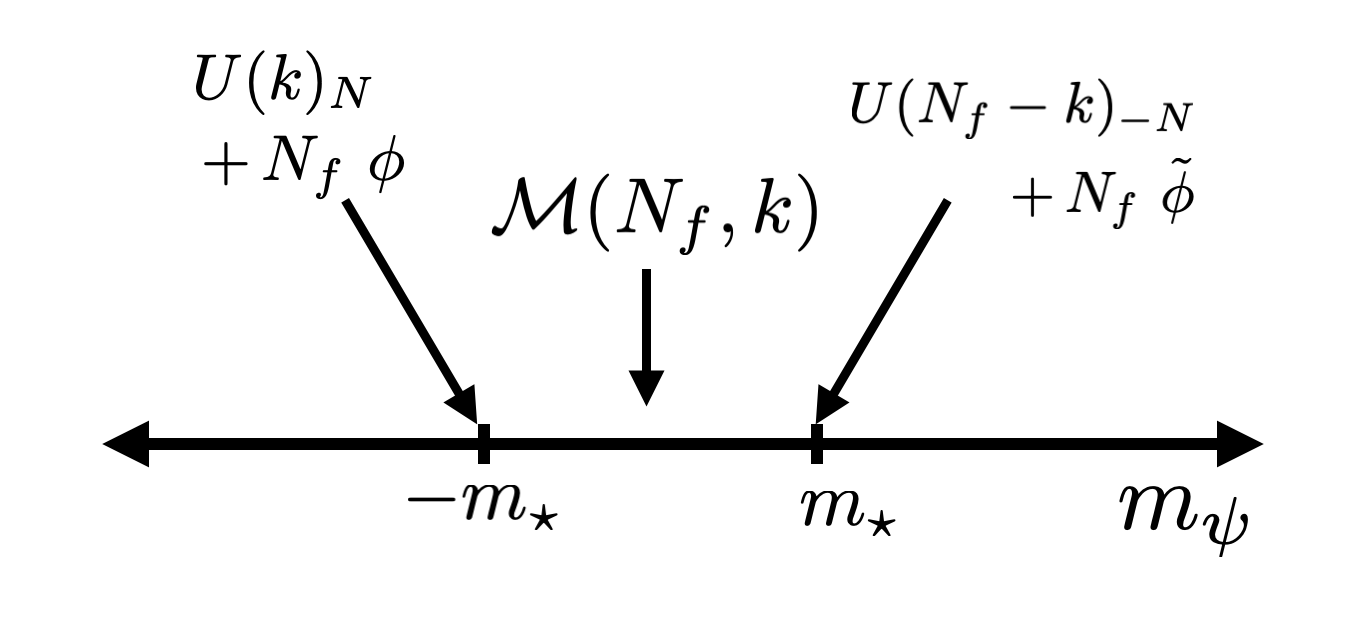}
\par\end{centering}
\caption{Symmetry breaking of QCD$_3$ according to \cite{Komargodski:2017keh}}\label{fig:breaking}
\end{figure}

At least this was the story until 2017 when it was conjectured that \eqref{eq:aharony} could be extended beyond the flavor bound. The authors of \cite{Komargodski:2017keh} proposed a scenario in which QCD$_3$ with non-zero Chern-Simons level actually confines for a finite range of the fermion mass, so long as $k<N_f<N_{\star}$ where $N_{\star}$ is a yet-to-be-determined function of $N$ and $k$. This may seem surprising since pure Chern-Simons theory is not confining for non-zero $k$, but nevertheless the conjecture passes many of the same consistency checks as \eqref{eq:aharony}. The novelty of this scenario is that the fermionic theory admits two scalar duals at finite value of the fermion mass. Specifically, we have the following duality \footnote{We can not say for certain that the critical points are symmetric about the origin, but for simplicity we will assume so. } 
\begin{equation}
SU(N)_{-k+N_{f}/2}\,\text{with }N_{f}\,\psi \, \, \, \, \,\leftrightarrow\, \, \, \, \,\begin{cases}
U(k)_{N}\,\text{with }N_{f}\,\phi & m_{\psi}=-m_{\star}\\
U(N_f-k)_{-N}\,\text{with }N_{f}\,\,\,\tilde{\phi} & m_{\psi}=m_{\star}
\end{cases}.\label{eq:flq aharony}
\end{equation}
Between these phases we have a non-linear sigma model phase described by the complex Grassmannian 
\begin{equation}\label{eq:grassman}
\mathcal{M}(N_f,k)=\frac{U(N_f)}{U(k)\times U(N_f-k)}
\end{equation}
supplemented by an appropriate Wess-Zumino term denoted by $\Gamma$. Since either scalar theory is only a valid dual for a finite patch of the phase diagram, the radius of the NLSM remains small and the theory can not be seen semi-classically. This can be thought of as the ``chiral Lagrangian" of  QCD$_3$--a quark bilinear spontaneously obtains a vev which breaks the flavor symmetry leading to the coset \eqref{eq:grassman}. These massless degrees of freedom are the mesons of the 2+1 d confining phase. A pictorial summary of this situation is given in Fig. \ref{fig:breaking}. 

We are interested in extending the analysis of \cite{Komargodski:2017keh} to incorporate the effects of unequal quark masses. Specifically, we explicitly break the flavor group $U(N_f)$ to $U(f)\times U(F)$ and then match the resulting theories onto \eqref{eq:aharony} or \eqref{eq:flq aharony}. We find a rich structure of Grassmannians which disappear and reemerge as we tune $F$ and $f$ relative to $k$. Our results reduce to that of \cite{Komargodski:2017keh} in the limit where both quark masses are equal. Our phase diagram as a function of $f$ and $F$ is given in Fig. \ref{fig:sum}. In section \ref{sec:phases} we lay out the phases diagrams as a function of mass for each relevant case. In section \ref{sec:scalar} we describe how to obtain the rich network of Grassmannians from the scalar theories and the necessary potentials needed to obtain this structure. We conclude in section \ref{sec:conclude}.

As this work was being finalized we learned of \cite{Argurio:2019tvw} which has overlap with this work. Luckily our results are in agreement.

\section{Constructing the Phase Diagram} \label{sec:phases}
We start with a flavor group of $U(F)\times U(f)$ with $F+f=N_{f}$. Without loss of generality, we take $f\le F$. We denote $F$ fermions by $\Psi$ with mass $M$ and the $f$ fermions by $\psi$ with mass $m$. We always work with ``bare" Chern-Simons level, explicitly displaying the contribution from the $\eta$ invariant. Moreover, $k$ will always be positive. A negative Chern-Simons level will always come with an explicit minus sign.
\subsection{Flavor Bounded Case} \label{sec:flav_bound}
\begin{figure}[t]
\begin{centering}
\includegraphics[scale=0.30]{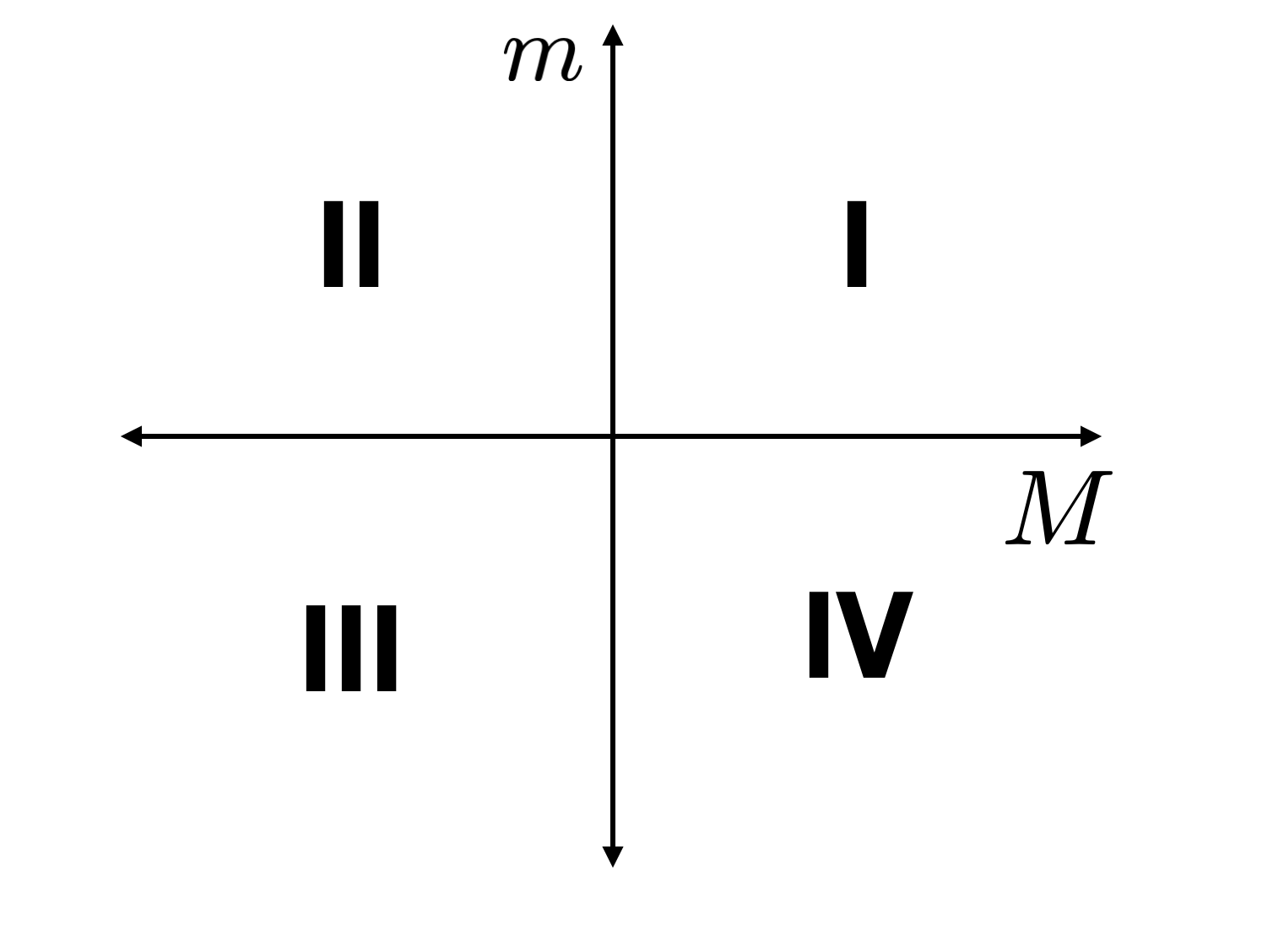}
\par\end{centering}
\caption{Phases when $f+F<k$}\label{fig:flav_bounded}
\end{figure}

We start by considering the $f+F\le k$ case for completeness. We obtain a phase diagram as in Fig. \ref{fig:flav_bounded}. Any value of the mass deformation leads to a theory which satisfies the flavor bound. As a result, we have four distinct topological field theories seperated by critial lines hosting light fermions. These lines admit bosonized duals in accordance with \cite{Aharony:2015mjs}. The completely gapped phases give the following TFTs:

\begin{subequations}
\begin{align}
\text{I)}:\qquad &  SU(N)_{-k+f+F} \text{     } \leftrightarrow \text{     }U(k-f-F)_N\\
\text{II)}:\qquad & SU(N)_{-k+f} \text{     } \leftrightarrow \text{     }U(k-f)_N\\
\text{III)}:\qquad & SU(N)_{-k} \text{     } \leftrightarrow \text{     }U(k)_N\\
\text{IV)}:\qquad & SU(N)_{-k+F} \text{     } \leftrightarrow \text{     }U(k-F)_N
\end{align}
\end{subequations}
while the critical lines host
\begin{subequations}
\begin{align}
\text{I-II)}:\qquad &  SU(N)_{-k+f+\frac{F}{2}} \text{ with } F \,\, \Psi \leftrightarrow U(k-f)_N \text{ with } F \,\, \Phi \\
\text{II-III)}:\qquad & SU(N)_{-k+\frac{f}{2}} \text{ with } f \, \, \psi \leftrightarrow \text{     } U(k)_N \text{ with } f \,\, \phi \\
\text{III-IV)}:\qquad &  SU(N)_{-k+\frac{F}{2}} \text{ with } F \,\, \Psi \leftrightarrow U(k)_N \text{ with } F \,\, \Phi \\
\text{IV-I)}:\qquad &  SU(N)_{-k+F+\frac{f}{2}} \text{ with } f \,\, \psi \leftrightarrow U(k-F)_N \text{ with } f \,\, \phi. 
\end{align}
\end{subequations}
This scenario is relatively straightforward since there are no quantum phases. More structure emerges as we allow violation of the flavor bound. There is also the added complication of a possible $U(f)\times U(F)$ potential on the scalar side. This will not drastically effect the structure of this phase diagram. See Sec. \ref{sec:scalar} for more details.

\subsection{Flavor Violated Case} \label{sec:flav_vi}
We now turn to the case where $k<N_f<N_{\star}$. Right away we note that if $m=M$ our flavor group is enchanced from $U(f)\times U(F)\to U(f+F)$ and we are back to the case considered in \cite{Komargodski:2017keh}. This will serve as an important benchmark for the complete phase diagram. 

For the case of two flavor groups we obtain six separate phase diagrams depending on the relative sizes of $f,F$ and $k$. These are 
\begin{enumerate}[label=\roman*]
\centering
 \item[i.)] $f<k$, $F<k$
\item[ii.)] $f<k$, $F>k$
\item[iii.)] $f>k$, $F>k$
\item[iv.)] $f<k$, $F=k$
\item[v.)] $f=k$, $F>k$
\item[vi.)] $f=k$,$F=k$.
\end{enumerate}
The procedure for mapping out these phase diagrams is identical in all six phases. We start by gapping out all of the matter to determine the asymptotic TFTs. Next, we keep one flavor light and give the other a positive or negative mass deformation. This reduces the problem to a single flavor group with a shifted Chern-Simons level. The resulting asymptotic theory will either satisfy the flavor bound with the new CS level and admit a single scalar dual or violate it and have a quantum phase with two scalar duals. We will assume that $0<k<f+F<N_{\star}$.

Before we proceed we would like to highlight one subtlety in the procedure above. When examining the flavor bounds we must take the effective CS level to have opposite sign as the shift coming from the $\eta$ invariant. This is the convention used throughout the literature and is necessary for an accurate application of the dualities. As an example say we encounter a situation where the CS level is $k+\frac{f}{2}$. The dualities in \cite{Aharony:2015mjs,Komargodski:2017keh} say nothing about this situation as written. To apply the dualities correctly, we shift $k$ such that we obtain $k+f -\frac{f}{2}$ and apply the flavor bounds with respect to $k_{eff}=k+f$. This simply corresponds to changing the sign of the $\eta$ invariant in the path integral \cite{Witten:2015aba}. This may look disturbing, since it seems as if we are using different regularizations in separate parts of the phase diagram. However this is not the case. We are simply shifting the levels of the induced CS terms in such a way that it appears as if we are changing the sign of the $\eta$ invariant, without actually explicitly performing the necessary time-reversal operation to do so. Stated another way--the partition function of an $SU(N)_{k}$ fermionic theory with a positive $\eta$-invariant is equivalent to an $SU(N)_{k+f}$ fermionic theory with a negative $\eta$-invariant\footnote{This is another reason why we choose to work with the bare CS level. This procedure is straightforward to implement in this case. It is obscured by working with the full quantum CS level.}. Let us now turn to the phase diagrams for these six cases.

\subsubsection{$f,F<k$}\label{sec:bothless}
\begin{figure}
\begin{centering}
\includegraphics[scale=0.35]{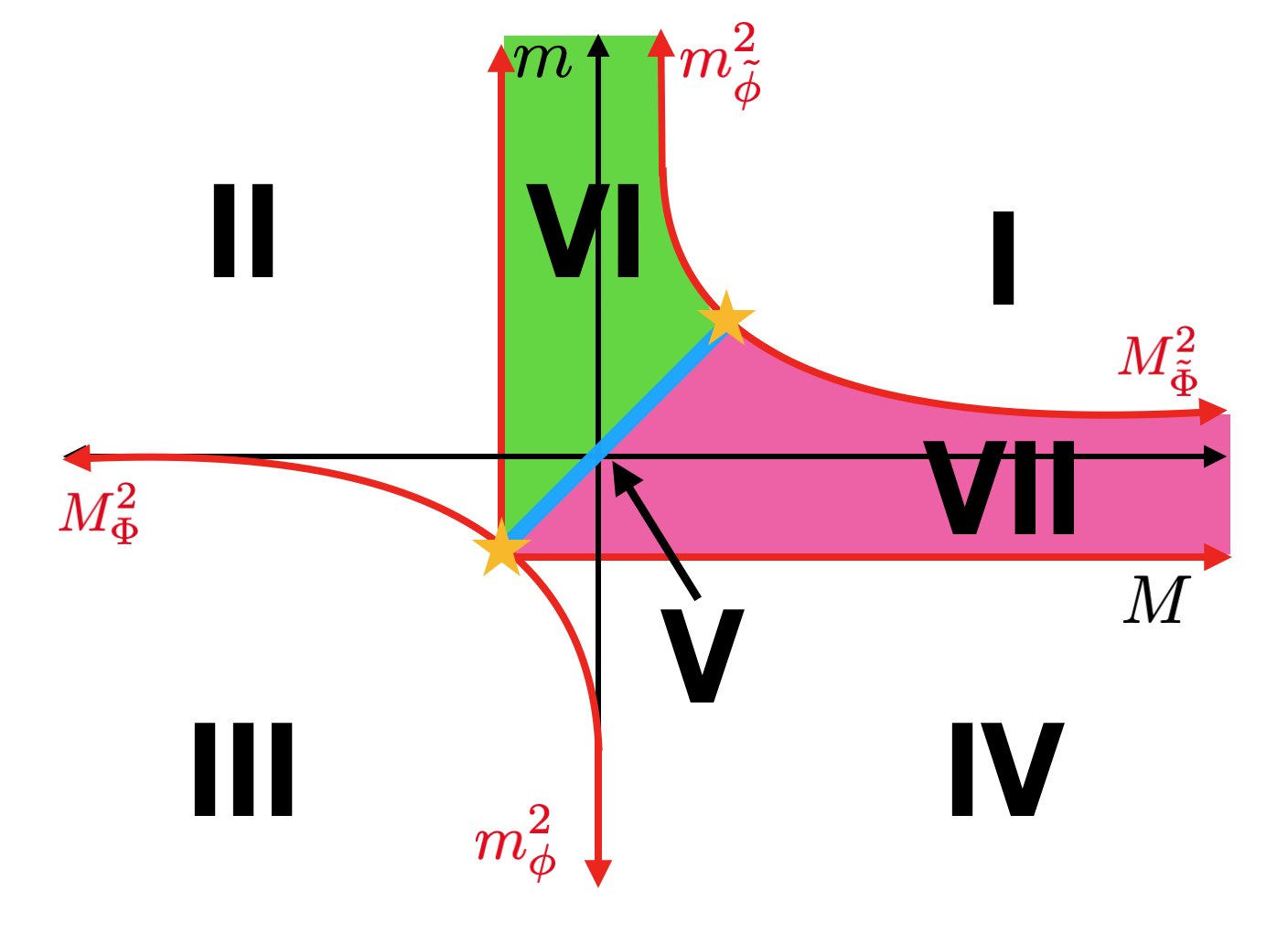}
\par\end{centering}
\caption{Phase diagram for flavor broken QCD$_3$ with $f,F<k$. The phases are described in Sec. \ref{sec:bothless}. The stars denote the bosonic duals given in \cite{Komargodski:2017keh}.}\label{fig:both_less}
\end{figure}

The phase diagram for this case is given in Fig. \ref{fig:both_less}. Explicit mass labels correspond to positive values of the mass deformation. Both flavors of fermions condense. Naively this may not seem possible since both groups of flavors satisfy the flavor bounds. But when compared to the effective CS level induced by integrating out the other group, they become flavor violating. The phases are 
\begin{subequations}
\begin{align}
\text{I)}:\qquad &  SU(N)_{-k+f+F} \text{     } \leftrightarrow \text{     }U(F+f-k)_{-N}\\
\text{II)}:\qquad & SU(N)_{-k+f} \text{     } \leftrightarrow \text{     }U(k-f)_N\\
\text{III)}:\qquad & SU(N)_{-k} \text{     } \leftrightarrow \text{     }U(k)_N\\
\text{IV)}:\qquad & SU(N)_{-k+F} \text{     } \leftrightarrow \text{     }U(k-F)_N\\
\text{V)}:\qquad & \mathcal{M}(f+F,k) +N\,\Gamma\\
\text{VI)}: \qquad & \mathcal{M}(F,k-f) +N\,\Gamma \\
\text{VII)}: \qquad & \mathcal{M}(f,k-F) +N\,\Gamma 
\end{align}
\end{subequations}
while the critical lines (red) host the bosonic theories
\begin{subequations}
\begin{align}
\text{I-VI)}: \qquad & U(F+f-k)_{-N} \text{ with } F \, \tilde{\Phi} \\
\text{VI-II)}:\qquad & U(k-f)_{N} \text{ with } F\, \Phi \\
\text{II-III)}:\qquad &  U(k)_{N} \text{ with } f\, \phi \\
\text{III-IV)}:\qquad &  U(k)_{N} \text{ with } F\, \Phi\\
\text{IV-VII)}:\qquad & U(k-F)_{N} \text{ with } f\, \phi\\
\text{VII-I)}:\qquad & U(F+f-k)_{-N} \text{ with } f\,\tilde{\phi}.
\end{align}
\end{subequations}
The theories sitting at the stars in the third and first quadrant are $U(k)_{N}$ with $f\, \phi \, \, +\, \, F\, \Phi$ and $U(F+f-k)_{-N}$ with $f\, \tilde{\phi} \, \, +\, \, F\,\tilde{\Phi}$ respectivley. These are the CFTs corresponding to the flavor enchanced case of \cite{Komargodski:2017keh} and will be present in all of the remaining diagrams. Also note that the II-III and III-IV critical lines are dual to massless fermionic theories, while the others are dual to a fermionic theory with a mass offset. As the number of flavors decreases this mass offset should decrease as well, leading to the curvature of the critical lines.

\subsubsection{$f<k,\,\,F>k$}\label{sec:oneless}
\begin{figure}
\begin{centering}
\includegraphics[scale=0.35]{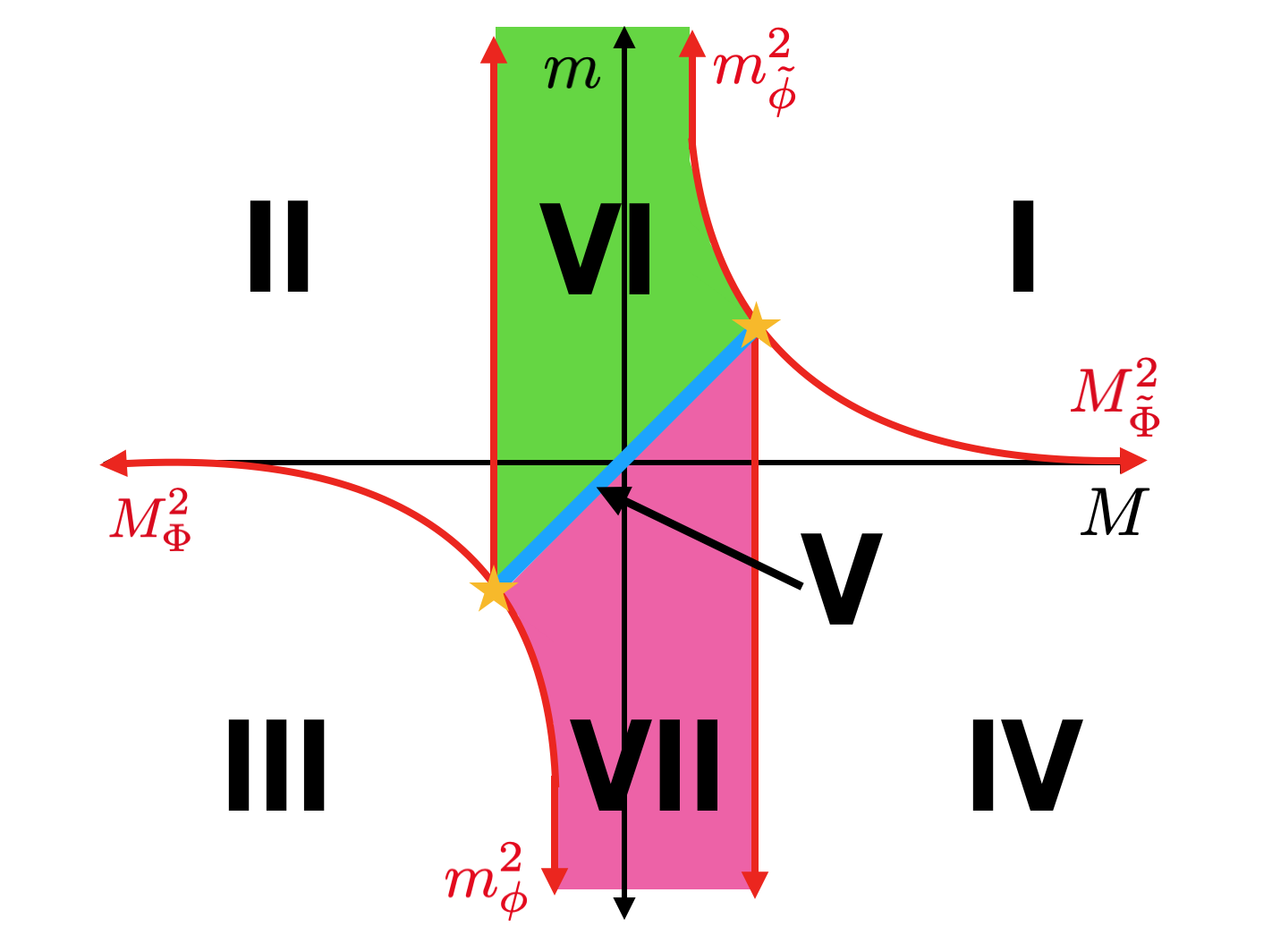}
\par\end{centering}
\caption{Phase diagram for flavor broken QCD$_3$ with $f<k,\,F>k$. These phases are laid out in Sec. \ref{sec:oneless}.}\label{fig:one_less}
\end{figure}

The phase diagram for this case is given in Fig. \ref{fig:one_less}. Here it is always the $F$ fermions which are condensing so long as we are away from the flavor enhanced line. The phases are 
\begin{subequations}\label{eq:one_more}
\begin{align}
\text{I)}:\qquad &  SU(N)_{-k+f+F} \text{     } \leftrightarrow \text{     }U(F+f-k)_{-N}\\
\text{II)}:\qquad & SU(N)_{-k+f} \text{     } \leftrightarrow \text{     }U(k-f)_N\\
\text{III)}:\qquad & SU(N)_{-k} \text{     } \leftrightarrow \text{     }U(k)_N\\
\text{IV)}:\qquad & SU(N)_{-k+F} \text{     } \leftrightarrow \text{     }U(F-k)_{-N}\\
\text{V)}:\qquad & \mathcal{M}(f+F,k) +N\,\Gamma\\
\text{VI)}: \qquad & \mathcal{M}(F,k-f) +N\,\Gamma \\
\text{VII)}: \qquad & \mathcal{M}(F,k) +N\,\Gamma 
\end{align}
\end{subequations}\label{eq:one_more_lines}
while the critical lines which separate them are given by the by the following bosonic CFTs:
\begin{subequations}
\begin{align}
\text{I-VI)}: \qquad & U(F+f-k)_{-N} \text{ with } F \, \tilde{\Phi} \\
\text{VI-II)}:\qquad & U(k-f)_{N} \text{ with } F\, \Phi \\
\text{II-III)}:\qquad &  U(k)_{N} \text{ with } f\, \phi \\
\text{III-VII)}:\qquad &  U(k)_{N} \text{ with } F\, \Phi\\
\text{VII-IV)}:\qquad & U(F-k)_{-N} \text{ with } F\, \tilde{\Phi}\\
\text{VI-I)}:\qquad & U(F+f-k)_{-N} \text{ with } f\,\tilde{\phi}.
\end{align}
\end{subequations}
Once again the theories sitting at the stars in the third and first quadrants are $U(k)_{N}$ with $f\, \phi \, \, +\, \, F\, \Phi$ and $U(F+f-k)_{-N}$ with $f\, \tilde{\phi} \, \, +\, \, F\,\tilde{\Phi}$ respectively.

\subsubsection{$f,F>k$}\label{sec:bothmore}
\begin{figure}[t]
\begin{centering}
\includegraphics[scale=0.35]{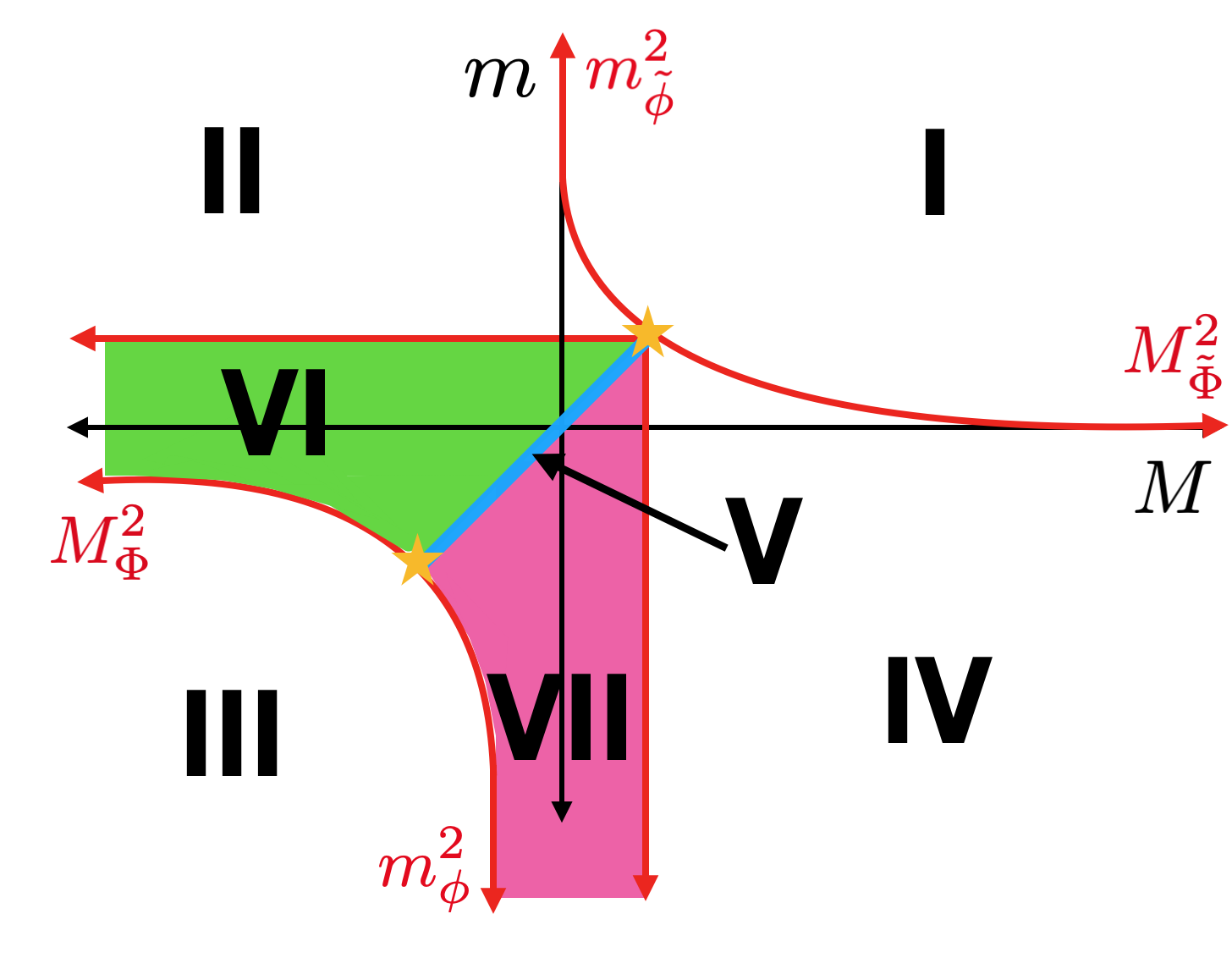}
\par\end{centering}
\caption{Phase diagram for flavor broken QCD$_3$ with $f,F>k$. These phases are laid out in Sec. \ref{sec:bothmore}.}\label{fig:both_more}
\end{figure}

The phase diagram for this case is given in Fig. \ref{fig:both_more}. Here both sets of fermions are condensing. The phases are 
\begin{subequations}\label{eq:both_more}
\begin{align}
\text{I)}:\qquad &  SU(N)_{-k+f+F} \text{     } \leftrightarrow \text{     }U(F+f-k)_{-N}\\
\text{II)}:\qquad & SU(N)_{-k+f} \text{     } \leftrightarrow \text{     }U(f-k)_{-N}\\
\text{III)}:\qquad & SU(N)_{-k} \text{     } \leftrightarrow \text{     }U(k)_N\\
\text{IV)}:\qquad & SU(N)_{-k+F} \text{     } \leftrightarrow \text{     }U(F-k)_{-N}\\
\text{V)}:\qquad & \mathcal{M}(f+F,k) +N\,\Gamma\\
\text{VI)}: \qquad & \mathcal{M}(f,k) +N\,\Gamma \\
\text{VII)}: \qquad & \mathcal{M}(F,k) +N\,\Gamma 
\end{align}
\end{subequations}\label{eq:both_more_lines}
while the critical lines which separate them are given by the by the following bosonic CFTs
\begin{subequations}
\begin{align}
\text{I-II)}: \qquad & U(F+f-k)_{-N} \text{ with } F \, \tilde{\Phi} \\
\text{II-VI)}:\qquad & U(f-k)_{N} \text{ with } f\, \tilde{\phi} \\
\text{VI-III)}:\qquad &  U(k)_{N} \text{ with } f\, \phi \\
\text{III-VII)}:\qquad &  U(k)_{N} \text{ with } F\, \Phi\\
\text{VII-IV)}:\qquad & U(F-k)_{-N} \text{ with } F\, \tilde{\Phi}\\
\text{VI-I)}:\qquad & U(F+f-k)_{-N} \text{ with } f\,\tilde{\phi}.
\end{align}
\end{subequations}
We note that this phase can be derived from the $f,F<k$ case of Sec. \ref{sec:bothless}  by a clever use of time reversal symmetry. This maps the theory with $f,F<k$ to one with $f,F>\tilde{k}$ where $\tilde{k}=F+f-k$. We then have $\mathcal{M}(F,f-k)=\mathcal{M}(F,\tilde{k})$. The action of time reversal also maps $m\to -m$ and so will change the location of this phase as well\footnote{We thank the authors of \cite{Argurio:2019tvw} for discussion on this mapping.}. This mapping was used in \cite{Argurio:2019tvw} to deduce the form of Fig. \ref{fig:both_more} without explicitly performing the mass deformations.

\subsubsection{$f<k,\, F=k$} \label{sec:oneeqless}
The case when either $f$ or $F$ is equal to $k$ offers a unique challenge. Consider the case at hand: $f<k$ and $F=k$. When we integrate out $F$, we will cancel the CS term completely and the Grassmannian corresponding to phase VII in Fig. \ref{fig:both_less} will disappear. As we increase $F$ further, the Grassmannian will reappear as phase VII in Fig. \ref{fig:one_less}. This offers an interesting view of the phase diagram as a function of $F$ and $f$. As we smoothly change $F$ through $k$, one of our Grassmannians disappears--we no longer have symmetry breaking in that part of the diagram. The theory is confining, all quarks pick up a mass from some confinement mechanism and the low energy theory is trivial. Moreover, in sharp distinction from the cases where neither $f$ nor $F$ is equal to $k$, we only have two Grassmannians instead of one. The phases corresponding to VI in Fig. \ref{fig:both_less} remains, however, since the effective CS level from integrating out the $f$ fermions ensures that the flavor bound is violated in this regime. The phases are:

\begin{subequations}\label{eq:one_eq_less}
\begin{align}
\text{I)}:\qquad &  SU(N)_{f} \text{     } \leftrightarrow \text{     }U(f)_{-N}\\
\text{II)}:\qquad & SU(N)_{-k+f} \text{     } \leftrightarrow \text{     }U(k-f)_{N}\\
\text{III)}:\qquad & SU(N)_{-k} \text{     } \leftrightarrow \text{     }U(k)_N\\
\text{IV)}:\qquad & SU(N)_{0}=\text{ trivial}\\
\text{V)}:\qquad & \mathcal{M}(f+F,k) +N\,\Gamma\\
\text{VI)}: \qquad & \mathcal{M}(k,k-f) +N\,\Gamma \\
\end{align}
\end{subequations}\label{eq:one_eq_less_lines}
The resulting CFTs on the critical lines are
\begin{subequations}
\begin{align}
\text{I-VI)}: \qquad & U(f)_{-N} \text{ with } k \, \tilde{\Phi} \\
\text{VI-II)}:\qquad & U(k-f)_{N} \text{ with } k\, \Phi \\
\text{II-III)}:\qquad &  U(k)_{N} \text{ with } f\, \phi \\
\text{III-VII)}:\qquad &  U(k)_{N} \text{ with } k\, \Phi\\
\text{IV-I)}:\qquad & U(f)_{-N} \text{ with } f\,\tilde{\phi}.
\end{align}
\end{subequations}
The III-IV and IV-I critical theories are consistent with the above since completely breaking the gauge group will lead to a trivial theory in the IR \cite{Benini:2017aed,Jensen:2017bjo}.

\begin{figure}[t]
\begin{centering}
\includegraphics[scale=0.35]{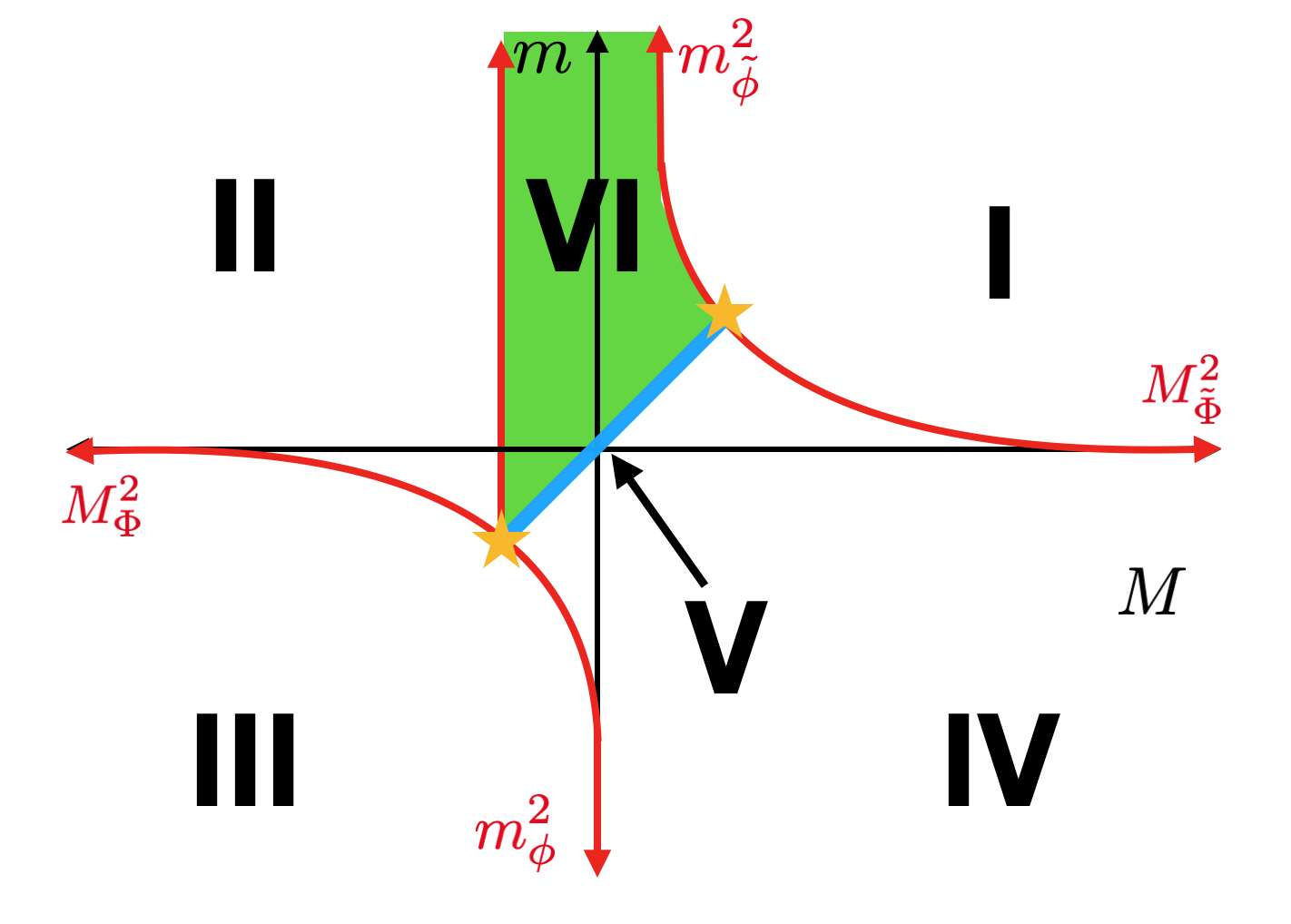}
\par\end{centering}
\caption{Phase diagram for flavor broken QCD$_3$ with $f<k,\, F=k$. These phases are described in Sec. \ref{sec:oneeqless}.}\label{fig:one_eq_less}
\end{figure}

\subsubsection{$f=k,\,F>k$}\label{sec:oneeqmore}
\begin{figure}[t]
\begin{centering}
\includegraphics[scale=0.35]{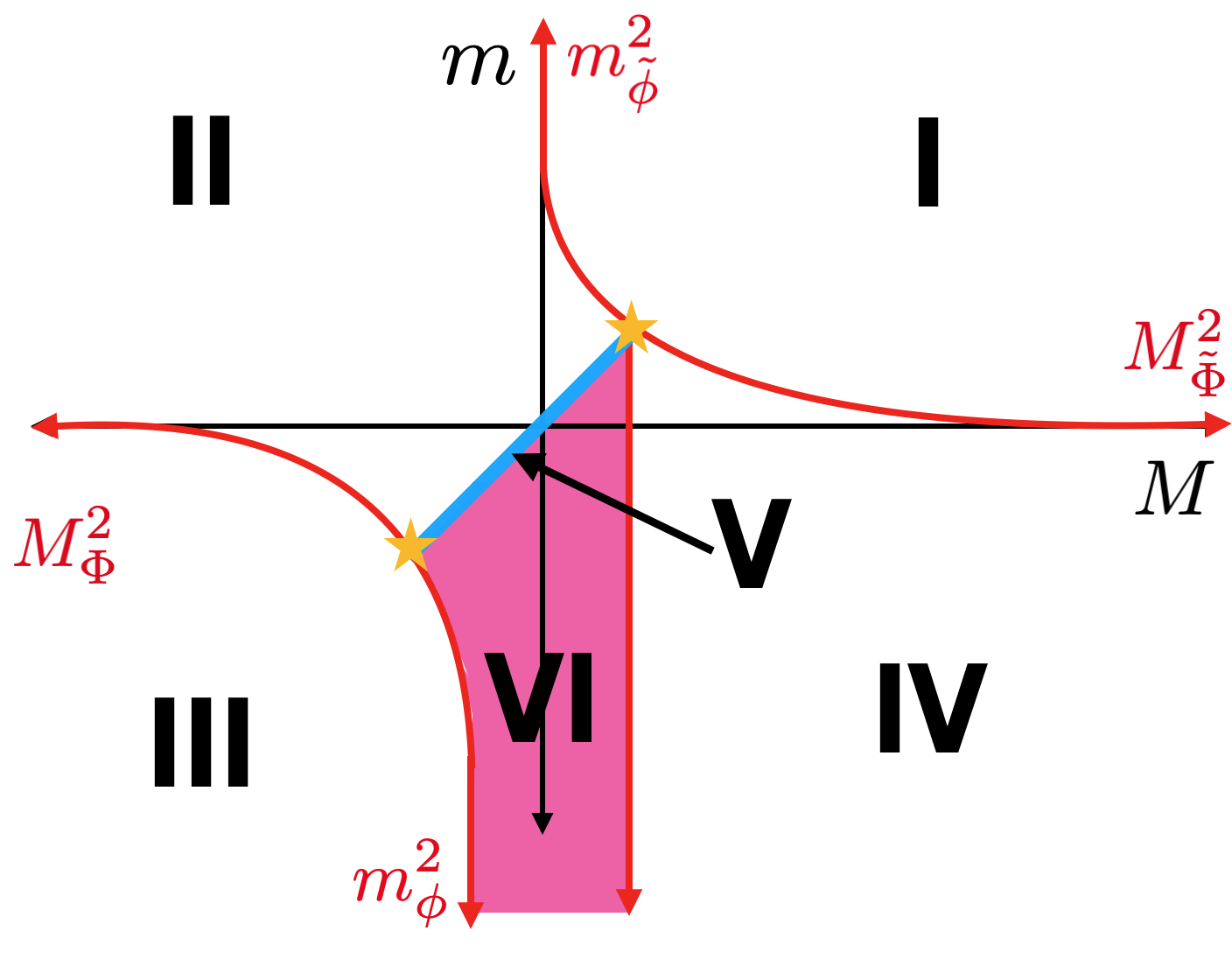}
\par\end{centering}
\caption{Phase diagram for flavor broken QCD$_3$ with $f=k,\, F>k$. These phases are laid out in Sec. \ref{sec:oneeqmore}.}\label{fig:bone_eq_more}
\end{figure}

This case is similar to the above in that one of the Grassmannian phases disappears while the other persists. This time, however, it is the Grassmannian which corresponds to phase VI in the $f<k,F>k$ (Fig. \ref{fig:one_less}) diagram which disappears. This has the same interpretation as the previous case--namely as we smoothly vary $f$ through $k$ the Grassmannian disappears and reappears as phase VI in the $f,F>k$ (Fig. \ref{fig:both_more}) phase diagram. At the transition we again only have two Grassmannians. The phase diagram is shown in Fig.\ref{fig:bone_eq_more}. The phases are

\begin{subequations}\label{eq:one_eq_less}
\begin{align}
\text{I)}:\qquad &  SU(N)_{F} \text{     } \leftrightarrow \text{     }U(F)_{-N}\\
\text{II)}:\qquad & SU(N)_{0} = \text{ trivial }\\
\text{III)}:\qquad & SU(N)_{-k} \text{     } \leftrightarrow \text{     }U(k)_N\\
\text{IV)}:\qquad & SU(N)_{F}=U(F)_{-N}\\
\text{V)}:\qquad & \mathcal{M}(f+F,k) +N\,\Gamma\\
\text{VI)}: \qquad & \mathcal{M}(F,k) +N\,\Gamma \\
\end{align}
\end{subequations}\label{eq:one_eq_less_lines}
while the bosonic CFTs are
\begin{subequations}
\begin{align}
\text{I-II)}: \qquad & U(F)_{-N} \text{ with } F \, \tilde{\Phi} \\
\text{II-III)}:\qquad & U(k)_{N} \text{ with } k\, \phi \\
\text{III-VI)}:\qquad &  U(k)_{N} \text{ with } F\, \Phi \\
\text{VI-IV)}:\qquad &  U(F-k)_{-N} \text{ with } k\, \tilde{\Phi}\\
\text{IV-I)}:\qquad & U(F)_{-N} \text{ with } k\,\tilde{\phi}.
\end{align}
\end{subequations}
The III-IV and IV-I critical theories are again consistent as explained in the previous subsection. This diagram can also be deduced from the action of time reversal on Fig. \ref{fig:bone_eq_more} as described at the end of Sec. \ref{sec:bothmore}. 

\subsubsection{$f=k,\,F=k$}\label{botheq}
\begin{figure}[t]
\begin{centering}
\includegraphics[scale=0.35]{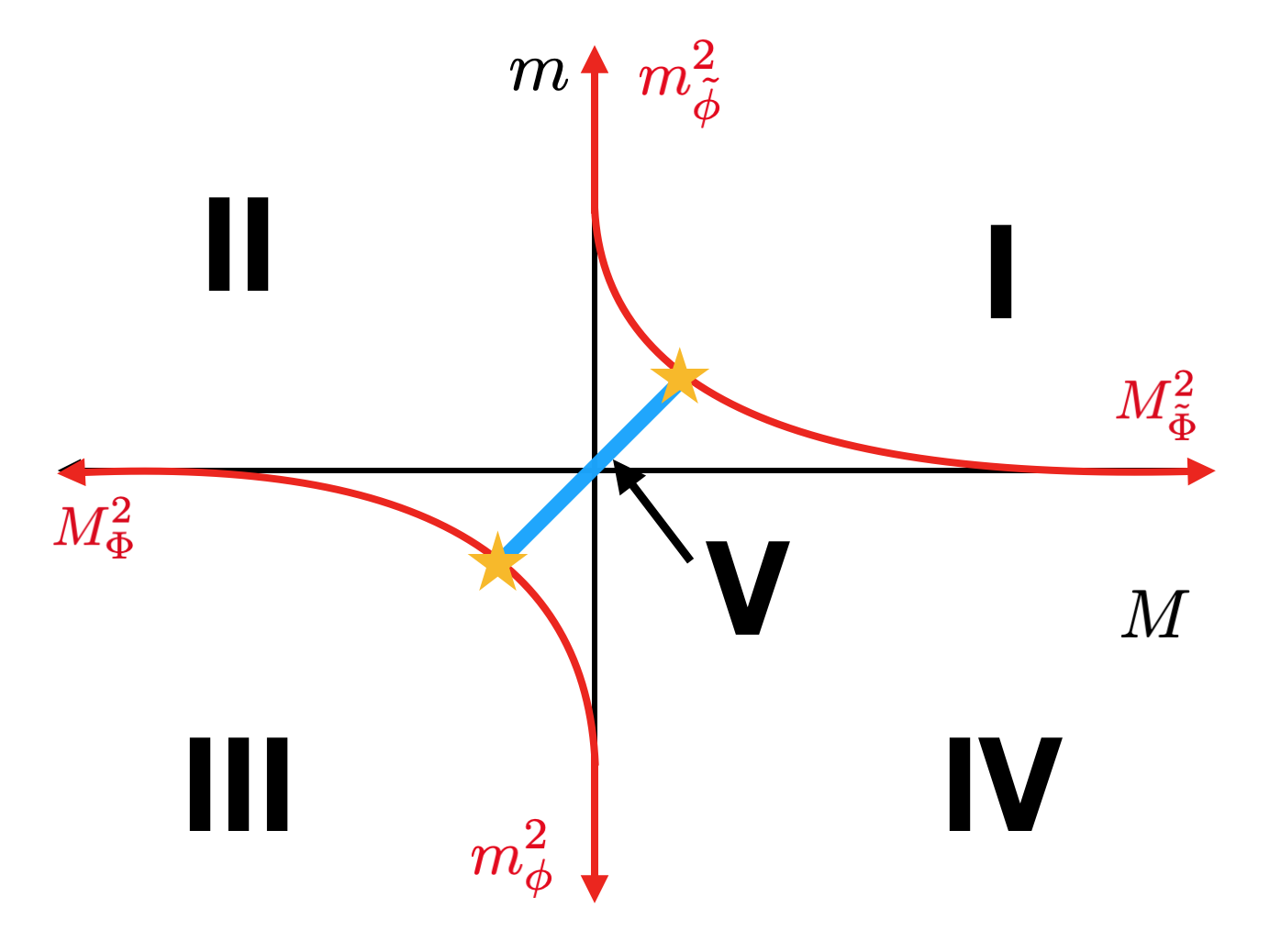}
\par\end{centering}
\caption{Phase diagram for flavor broken QCD$_3$ with $f=k,\, F=k$. These pahses are described in Sec. \ref{botheq}. This symmetry breaking pattern was first conjectured by Vafa and Witten \cite{Vafa:1984xh}.}\label{fig:both_eq}
\end{figure}

Finally we consider the case when both flavor are equal to $k$. In this case, we only get one Grassmannian phase living on the flavor-enhanced diagonal with two trivially gapped theories on either side. This case is given in \ref{fig:both_eq}
\begin{subequations}\label{eq:one_eq_less}
\begin{align}
\text{I)}:\qquad &  SU(N)_{k} \text{     } \leftrightarrow \text{     }U(k)_{-N}\\
\text{II)}:\qquad & SU(N)_{0} = \text{ trivial }\\
\text{III)}:\qquad & SU(N)_{-k} \text{     } \leftrightarrow \text{     }U(k)_N\\
\text{IV)}:\qquad & SU(N)_{0}=\text{ trivial }\\
\text{V)}:\qquad & \mathcal{M}(2k,k) +N\,\Gamma\\
\end{align}
\end{subequations}\label{eq:one_eq_less_lines}
while the bosonic CFTs are
\begin{subequations}
\begin{align}
\text{I-II)}: \qquad & U(k)_{-N} \text{ with } k \, \tilde{\Phi} \\
\text{II-III)}:\qquad & U(k)_{N} \text{ with } k\, \phi \\
\text{III-IV)}:\qquad &  U(k)_{N} \text{ with } k\, \Phi \\
\text{IV-I)}:\qquad & U(k)_{-N} \text{ with } k\,\tilde{\phi}.
\end{align}
\end{subequations}
This is nothing more than the flavor broken generalization of the symmetry breaking put forward by Vafa and Witten \cite{Vafa:1984xh}. The Vafa-Witten theorem then says that there can be no further symmetry breaking than what occurs on the flavor-enhanced diagonal. It is a reassuring check that our methods satisfy this theorem.

\subsection{Adding More Flavor Groups}
The next obvious generalization is to give a different mass to a third subset of flavors. This would further break the flavor symmetry to $U(N_f)\to U(f_1)\times U(f_2)\times U(f_3)$. The resulting phase diagram as a function of mass is now three dimensional so we will not try to reproduce it here. However we do have a few comments. Giving equal masses to each flavor will again reduce the problem to the single flavored case considered in \cite{Komargodski:2017keh}. These will lie on the line given by $m_1=m_2=m_3$. Now giving an asymptotically large mass to one of the flavors will reduce to one of the cases considered in Sec. \ref{sec:flav_vi}. The resulting scalar theories in that part of the phase diagram will not be localized on a curve as in Sec. \ref{sec:flav_vi}, but on a curved plane embedded in the three dimensional diagram. In between these planes will be the appropriate Grassmannians, if the effective CS level is flavor violating. There will now be a web of Grassmannians of differing dimensions permeating the diagram. As we tune any of the $f_i$ to $k$, then a corresponding Grassmannian will disappear and one of the TFTs will become trivial, as in the 2-flavor case. In general, the scalar CFTs will be localized on co-dimension 1 surfaces and there will be a plethora of flavor enhanced critical theories of various codimensionality where the different flavor groups become mass degenerate.

The above procedure is difficult to implement in practice, however. That is because the existence of flavor-violated phases not only relies on the relative sizes of the $f_i$'s, but the size of the pairwise sum of $f_i$'s with respect to $k$ . For example, consider a situation where $f_1<k$ with $f_1<f_2<f_3$. The two-flavor theory we are left with is then $SU(N)_{-k+f_1+\frac{f_2+f_3}{2}}$ with $f_2 \, \psi_2$ and $f_3 \, \psi_3$. Now give $f_2$ a negative mass. The existence of a quantum phase then depends on if $f_3>k-f_1$ or $f_3+f_1>k$. The amount of data needed to map out the phase diagram grows exponentially with the addition of each successive flavor group. Nevertheless this procedure can be implemented inductively as a way to map out the phase diagram as a function of each fermion mass individually.

\subsection{Summary}
\begin{figure}[t]
\begin{centering}
\includegraphics[scale=0.40]{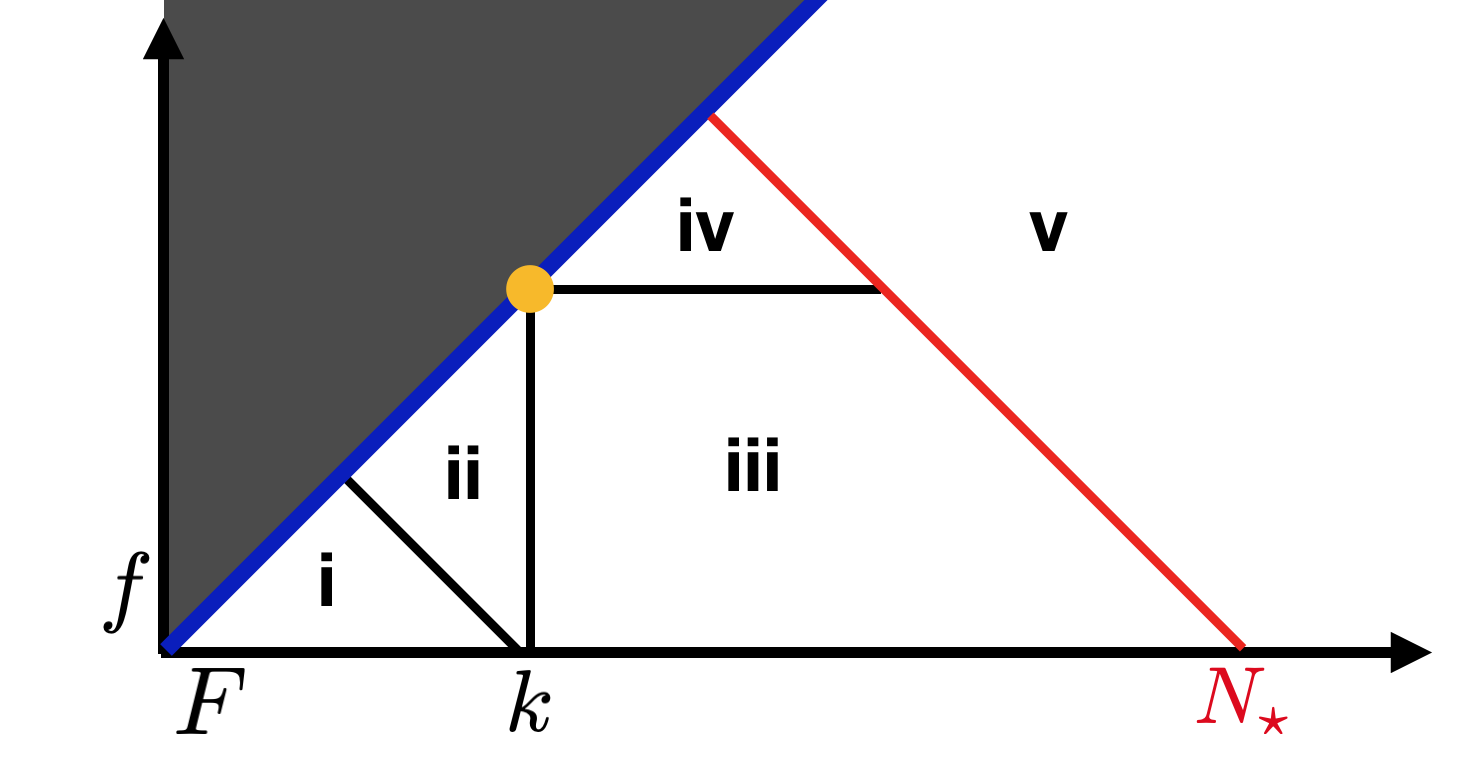}
\par\end{centering}
\caption{Phases as a function of $f$ and $F$.}\label{fig:sum}
\end{figure}

If our conjectured phases diagram is correct, the phases of QCD$_3$ as a function of $f$ and $F$ are given in Fig. \ref{fig:sum}. The black excision from the top right reflects the fact that we have restricted our analysis to $f\le F$. We can simply reflect the diagram about the diagonal if we are so inclined but the pertinant information is displayed. The solid blue line is again a reduction to the single flavor analysis considered in \cite{Komargodski:2017keh}. The various regions are:
\begin{enumerate}[label=\roman*]
\item[(i):] This corresponds to the $f+F<k$ considered in \ref{sec:flav_bound}. There are no quantum phases, just four distinct TFTs separated by CFTs with a single scalar duals.
\item[(ii):] This corresponds to the $f,F<k$ case considered in Sec. \ref{sec:bothless}. There are three Grassmannians and four TFTs separated by six scalar CFTs.
\item[(iii):] This corresponds to the $f<k,\, F>k$ case considered in Sec. \ref{sec:oneless}. There are again are three Grassmannians and four TFTs seperate by six scalar CFTs.
\item[(iv):] This corresponds to the $f,F>k$ case considered in Sec. \ref{sec:bothmore}. There are three Grassmannians and four TFTs seperate by six scalar CFTs 
\item[(v):] This corresponds to $N_f>N_{\star}$. There are again four distinct TFTs but no scalar CFTs seperating them.
\end{enumerate}
Sitting on the lines between these phases are
\begin{enumerate}
\item[(i)-(ii):] This transition marks the beginning of the existence of quantum phases. This line, however, is still captured by the duality in eq. \ref{eq:aharony}.
\item[(ii)-(iii):] This transition corresponds to $f<k,F=k$ case considered in Sec. \ref{sec:oneeqless}. Here there are three non-trivial TFTs, one trivial phase and two Grassmannians.
\item[(iii)-(iv):] This transition corresponds to $f=k,F<k$ case considered in Sec. \ref{sec:oneeqmore}. Again there are three non-trivial TFTs, one trivial phase and two Grassmannians.
\end{enumerate}
Finally the yellow dot corresponds to the Vafa-Witten breaking scenario \cite{Vafa:1984xh}. This lies on the $f=F$ line and is the crux of the entire phase diagram. Indeed, one can derive this entire symemtry breaking scenario from this breaking pattern alone \cite{Komargodski:2017keh}. 

The emergence, disappearance, and reemergence of various Grassmannians is an incredibly interesting phenomena. We believe this is indication that the phase transition between these phases is first order--the topology of the phase diagram as a function of $m$ and $M$ changes wildly when you tune $f$ and $F$. This must be taken with a grain of salt since $f$ and $F$ are not continuous parameters and so this is not a bonafide thermodynamic-type phase transition. It is possible that if we analytically continue $f$ and $F$ that the Grassmannians smoothly disappear and reappear as we tune, which would be indication of a second order ``phase" transition. The exact structure of the singly-saturated cases considered in Secs. \ref{sec:oneeqless} and \ref{sec:oneeqmore} indicate that a smooth transition may be possible. However Grassmannian manifolds with continuous dimensional is not a well-studied subject. Perhaps the answer is hiding in this complicated math. We leave a definitive answer to this important question for future work.  

\section{Comments on Scalar Potentials} \label{sec:scalar}
Let us briefly comment on the consistency of the scalar theories in our phase diagram. We use the following intuitive picture to guide our construction of the phase diagram: as we follow the RG flow from the UV to the IR, whichever mass deformation we encounter first will dictate the resulting dynamics. For example if we encounter a mass deformation which completely breaks the gauge group, this is will be the first thing to happen. The rest of the story is in the details of the scalar potential. With this picture in mind we can accurately say which scalars condense first in various parts of the phase diagrams. 

As with any discussion involving 3d bosonization dualities we are dealing with Wilson-Fisher scalars with a quartic potential tuned to the fixed point. As was recently shown in \cite{Argurio:2019tvw} and \cite{Armoni:2019lgb} this potential includes both single and double trace potentials of the form 
\begin{equation}\label{eq:potential}
\text{Tr} \left( \bar{\phi} \phi\right)^2 \text{  and   } \text{Tr} \left( (\bar{\phi} \phi)^2 \right).
\end{equation}
This potential is necessary to generate the maximal Higgsing pattern that has been used throughout the literature \footnote{The large $N$ analysis of \citep{Armoni:2019lgb} indicates that it may be necessary to include sextet terms as well, although it is possible that at finite $N$ the structure simplifies and only the quartics are necessary.}.  See the aforementioned references for a detailed discussion on how this occurs.

In addition to the quartics \eqref{eq:potential}, we will also assume that explicitly breaking the flavor symmetry in the UV will generate a certain $U(f)\times U(F)$ invariant relevant operator as we flow to the IR. This operator has the form 
\begin{equation} \label{eq:operator}
\mathcal{O}_4=c\,\phi^{\dagger}_{a\, I}\Phi^{a\, J}\Phi^{\dagger}_{b\, J} \phi^{b\, I}
\end{equation}
where $a,\, b$ are $SU(N)$ gauge indicies, $I$ is a $U(f)$ flavor index and $J$ is an $U(F)$ flavor index. This operator should be mirrored on the fermionic side as well, although it is not clear what it should be mapped to (for hints on how to do this see appendix A of \cite{Aitken:2018cvh}). For our conjecture to work, we need $c$ to be large and positive. This is similar in spirit to the quartic potential used in the master duality \cite{Benini:2017aed,Jensen:2017bjo} where it was crucial for the correct mapping of the phases. Indeed it performs the same task--when one scalar gets a vev, the other gains a large positive mass deformation for some of the gauge components. In analogy to \cite{Benini:2017aed,Jensen:2017bjo} we call the gauge components which gain a mass from \eqref{eq:operator} ``singlet scalars" since they are charged under the broken part of the gauge group. As an example consider the case when $f<k,F>0$ given in Fig. \ref{fig:one_less}. Assuming maximal Higgsing (which we will continue to do from here on out) as we give $\phi$ a negative mass deformation, the gauge group gets broken down to $U(k-f)\times U(f)$ and $\phi$ picks a vev which can be written as
\begin{equation}\label{eq:f_vev}
\langle \phi^{\dagger}_{a\,I} \phi_{b\, I} \rangle =
v\left( \begin{array}{cc}
I_{f\times f} & 0\\
0 & 0 
\end{array} \right)^{a\, b}
\end{equation}
This causes first $f$ gauge components of the $\Phi$ scalars to decouple from the unbroken part of the gauge group and obtain a large positive mass. The remaining $k-f$ light components can then drive the unbroken gauge group into the Grassmannian phase. If we considered the case where $f>k$ as well, then all of the components of $\Phi$ would get a large positive mass and would not contribute to the low energy dynamics, such as in the third quadrant of Fig. \ref{fig:both_more}. When both have the same mass, this does not occur and we again back to the case of \cite{Komargodski:2017keh}. This mechanism nicely explains the Grassmannian structures in Sec. \ref{sec:flav_vi} and why only one flavor can condense at a time independent of the presence of the single trace quartics in \eqref{eq:potential}. For consistency we must assume that once this potential is triggered along the RG flow, it no longer has any effect on the physics. That is the action of this potential is unidirectional with respect to whichever scalar get it's vev first. If $\phi$ gets triggered first, and then $\Phi$, the $\phi$ vev will give masses to the components of $\Phi$ but not vice versa.

Further, there is a possibility that this mass deformation for the singlet scalars can be cancelled off by an explicit negative mass deformation of the $\Phi$'s. This would result in extra lines of light singlet scalars within the Grassmannian phases that are decoupled from the dynamics (for a related discussion see \cite{TBD}). We assume that this is not the case by letting $c$ be larger than any other scale in the problem. This is, of course, rather artificial. Certainly this coefficient will flow to a specific value along the RG flow. What we are actually doing here is scaling one of the scalar fields relative to the other so that we zoom in to the part of the phase diagram where these interactions dominate \footnote{We thank Andreas Karch for discussion on this point.}.

In addition to the induced mass brought on by \eqref{eq:operator}, the potential \eqref{eq:potential} can also induce a mass deformation for one of the scalars. For example the double trace term will decompose as 
\begin{equation} \label{eq:potential}
\left( |\phi|^2+|\Phi|^2\right)^2=|\phi|^4+|\Phi|^4+2|\phi|^2|\Phi|^2
\end{equation}
Again we get that if one scalar condenses we again have $\langle \phi \rangle \ne 0$ the other flavor gets and induced mass deformation from the cross term in \eqref{eq:potential}. This can be compensated for by a suitable shift in the mass of $\Phi$ and so should only serve to shift the location of where $\Phi$ goes light in the phase diagram. This is different from the case discussed above because it is not various subsets of the gauge components which obtain a mass from this interaction. In other words this interaction does not lead to the existence of singlet scalars. We assume our red scalar theories in Sec. \ref{sec:flav_vi} take this effect into account. 

\section{Conclusion} \label{sec:conclude}

In this work we have mapped out the phases of QCD$_3$ with a product flavor group of the form $U(f)\times U(F)$. We have found a rich structure which involves multiple complex Grassmannians in different regions of the $(m,M)$ plane bounded by various Wilson-Fisher scalar theories. As $f$ and $F$ are tuned through $k$, some of these Grassmanians disappear and reappear in a different part of the $(m,M)$ plane. We also showed how a certain $U(f)\times U(F)$ invariant potential, which is expected to be generated along the RG flow, leads to this rich structure on the scalar side. This discussion can be straightforwardly generalized to symplectic and orthogonal gauge groups, since their breaking pattern is qualitatively the same, albeit with different Grassmannians.

Recent work on the ``standard" large N limit of QCD$_3$ has revealed an even richer quantum phase structure than previously imagined \citep{Armoni:2019lgb}. In short, the phase transitions between TFTs and quantum phases gets resolved into a series of first order phase transitions as you dial the mass. It would be interesting to see the emergence and reemergence of various Grassmannians in this limit and see whether these transitions still occur or not. 

One glaring unanswered question is how exactly we can see the transition between Grassmannians in the NLSM phase without mentioning the scalars. Indeed it should be the case that fermion mass deformations can be incorporated by adding a mass term for the fundamental fields of the Grassmannians, as was proposed in \cite{Komargodski:2017keh}. Upon first attempt this procedure looked promising, but it lacked a crucial ingredient unique to 2+1 d fermionic theories: the difference in positive and negative mass deformations for the fermions and their effect on the Chern-Simons level. Perhaps this point reflects the author's ignorance on the finer points of Grassmannian geometry. Nevertheless this is something that would be interesting to work out in all gory detail. 

There are still a lot of interesting questions to be answered, most pressing is the actual order of the phase transitions. We believe that this can be explored by a large $N,k,N_f$ analysis or by studying the structure of the Grassmannians as a continuous function of the dimension. In addition, perhaps this work can be extended to the case of $U(N)_k$ with fermions by gauging a global $U(1)$ symmetry and bringing it to the other side of \eqref{eq:aharony}. This will surely change the form of the Grassmannian, but it may allow us to study the duality in the Abelian limit where calculations are slightly easier. We could also imagine coupling our QCD$_3$ to scalar matter as in \cite{Benini:2017aed,Jensen:2017bjo}. Another interesting question is the lack of double condensation in the Grassmannian regime. While this question has been explored in detail in \cite{Argurio:2019tvw} using the gauged sigma model, it would be interesting to explore this from the fermionic perspective in spirit to the analysis of \cite{Vafa:1984xh}. In particular, is there a Vafa-Witten like theorem for non-zero Chern-Simons level?

\section*{Acknowledgments}
We would like to thank Andreas Karch for his invaluable advice in the preparation of this note and the authors of \citep{Argurio:2019tvw} for enlightening discussions related to this work. We also thank Kyle Aitken and Changha Choi for collaboration on related work. This work was supported, in part, by the U.S.~Department of Energy under Grant No.~DE-SC0011637.
\bibliographystyle{JHEP}
\bibliography{flavorbroken}
\end{document}